\documentclass[a4paper,12pt]{article}
\usepackage{amssymb}
\usepackage{latexsym}
\usepackage[dvips]{graphicx}
\usepackage{cite}
\newcommand{\be}{\begin{equation}}
\newcommand{\ee}{\end{equation}}
\newcommand{\bea}{\begin{eqnarray}}
\newcommand{\nn}{\nonumber}
\newcommand{\eea}{\end{eqnarray}}

\begin{document}

%\begin{titlepage}
\begin{flushright}
UB-ECM-PF-04/11
\end{flushright}
\begin{centering}
\vspace{.3in}
{\Large {\bf Energy distribution in a BTZ black hole spacetime}}
\\

\vspace{.5in} {\bf  Elias C.
Vagenas\footnote{evagenas@ecm.ub.es} }\\

\vspace{0.3in}

Departament d'Estructura i Constituents de la Mat\`{e}ria\\
and\\ CER for Astrophysics, Particle Physics and Cosmology\\
Universitat de Barcelona\\
Av. Diagonal 647\\ E-08028 Barcelona\\
Spain\\
\end{centering}

\vspace{0.7in}
%%%%%%%%%%%%%%%%%%%ABSTRACT%%%%%%%%%%%%%%%%%%%%%%%%%%%%%%%%%%
\begin{abstract}
We evaluate the energy distribution associated with the ($2+1$)-dimensional rotating BTZ black hole. The
energy-momentum complexes of Landau-Lifshitz and Weinberg are employed for this computation. Both prescriptions
give exactly the same form of energy distribution. Therefore, these results provide evidence in support of the
claim that, for a given gravitational background, different energy-momentum complexes
can give identical results in three dimensions, as it is the case in four dimensions.\\
 \end{abstract}

%%%%%%%%%%%%%%%%%%%%%%%%%%%%%%%%%%%%%%%%
%\end{titlepage}
\newpage

\baselineskip=18pt
%%%%%%%%%%%%%%%%%%%%%%%%%%%%%%%%%%%%%%%%%%%%%%%%%%%%%%%%%%%%%%%%%%%%%%%%%%%%%%%%%%%%%%%%%%%%%%%%%
%%%%%%%%%%%%%%%%%%%%%%%%%%%%%%%%%%%%%%%%%%%%%%%%%%%%%%%%%%%%%%%%%%%%%%%%%%%%%%%%%%%%%%%%%%%%%%%%
%%%%%%%%%%%%%%%%%%%%%%%%%%%%%%%%%%%%%%%%%%%%%%%%%%%%%%%%%%%%%%%%%%%%%%%%%%%%%%%%%%%%%%%%%%%%%%%%%
%%%%%%%%%%%%%%%%%%%%%%%%%%% INTRODUCTION %%%%%%%%%%%%%%%%%%%%%%%%%%%%%%%%%%%%%%%%%%%%%%%%%%%%%%%%
\section*{Introduction}
One of the oldest problems in gravitation which still lacks of a definite answer is the localization of energy and
momentum. Much attention has been devoted for this problematic issue. Einstein was the first to construct a
locally conserved energy-momentum complex \cite{einstein}. Consequently, a plethora of different energy-momentum
complexes  were proposed \cite{ll}-\hspace{-0.1ex}\cite{weinberg}.
These expressions were restricted to evaluate
energy distribution in quasi-Cartesian coordinates. M{\o}ller \cite{moller} proposed a new expression for an
energy-momentum complex which could be utilized to any coordinate system. However, the idea of the energy-momentum
complex was severely criticized for a number of reasons. Firstly, although a symmetric and locally conserved
object, its nature is nontensorial and thus its physical interpretation seemed obscure  \cite{chandra}. Secondly,
different energy-momentum complexes could yield different energy distributions for the same gravitational
background \cite{bergqvist1,bergqvist2}. Thirdly, energy-momentum complexes were local objects while there was
commonly believed that the proper energy-momentum of the gravitational field was only total, i.e. it cannot be
localized \cite{chiang}. For a long time, attempts to deal with this problematic issue were made only by proposers
of quasi-local approach \cite{brown,sean}.

\par\noindent
In 1990 Virbhadra revived the interest in this approach \cite{virbhadra}. At the same time Bondi \cite{bondi}
sustained that a nonlocalizable form of energy is not admissible in relativity so its location can in principle be
found. Since then, numerous works on evaluating the energy distribution of several gravitational backgrounds have
been completed employing the abandoned for a long time approach of energy-momentum complexes \cite{complexes}.

\par\noindent
In 1996 Aguirregabiria, Chamorro and Virbhadra \cite{virbhadra1} showed that five different\footnote{Later on
Virbhadra \cite{virbhadra2} came  to know that Tolman's and Einstein's complexes which had been used in
\cite{virbhadra1} were exactly the same (see footnote 1 in \cite{virbhadra2}).} energy-momentum complexes yield
the same energy distribution for any Kerr-Schild class metric. Additionally, their results were identical with the
results of Penrose \cite{pen} and Tod \cite{tod} using the notion of quasi-local mass.

\par\noindent
Later attempts to deal with this problematic issue were made (as already mentioned) by proposers of quasi-local
approach. The determination as well as the computation of the quasilocal energy and quasilocal angular momentum of
a ($2+1$)-dimensional gravitational background were first presented by Brown, Creighton and Mann \cite{mann1}.
Many attempts since then have been performed to give new definitions of quasilocal energy in General Relativity
\cite{quasilocal}. Considerable efforts  have also been performed in constructing superenergy tensors
\cite{senovilla1}. Motivated by the works of Bel \cite{bel} and independently of Robinson \cite{robinson}, many
investigations have been carried out in this field \cite{senovilla2}.

\par\noindent
In 1999 Chang, Nester and Chen \cite{nester} proved that every energy-momentum complex is associated with a
Hamiltonian boundary term. Thus, the energy-momentum complexes are quasi-local and acceptable.

\par\noindent
In this work the approach of energy-momentum complexes is implemented. The gravitational background  under
investigation is the ($2+1$)-dimensional rotating BTZ black hole \cite{jorge}. We evaluate the energy confined in
a ``one-sphere'' ($S^{1}$) of radius $r_0$ associated with the aforesaid background. Specifically, we are
implementing the prescriptions of Landau-Lifshitz and Weinberg. The specific ($2+1$)-dimensional black hole
background is described by two parameters: the mass $M$ and the angular momentum (spin) $J$. It is locally Anti-de
Sitter space and thus it differs from Schwarzschild and Kerr solutions in that it is asymptotically Anti-de Sitter
instead of flat. Additionally, it has no curvature singularity at the origin.
\par\noindent
The rest of the paper is organized as follows. In the next section we consider the concept of energy-momentum
complexes in the context of General Theory of Relativity. In Section 2 we briefly present the $(2+1)$-dimensional
BTZ black hole in which the energy  distribution by using two different prescriptions is to be calculated. In the
subsequent two sections using Landau-Lifshitz and Weinberg energy-momentum complexes, respectively, we explicitly
compute the energy distribution contained in a ``one-sphere" of fixed radius $r_0$. The results extracted from
these two different prescriptions associated with the same gravitational background are identical.
 Finally, in Section 5 a brief summary of results and concluding remarks are presented.
 %%%%%%%%%%%%%%%%%%%%%%%%%%%%%%%%%%%%%%%%%%%%%%%%%%%%%%%%%%%%%%%%%%%%%%%%%%%%%%%%%%%%%%%%%%%%%%%%%
%%%%%%%%%%%%%%%%%%%%%%%%%%%%%%%%%%%%%%%%%%%%%%%%%%%%%%%%%%%%%%%%%%%%%%%%%%%%%%%%%%%%%%%%%%%%%%%%%
%%%%%%%%%%%%%%%%%%%%%%%%%%%%%%%%%%%%%%%%%%%%%%%%%%%%%%%%%%%%%%%%%%%%%%%%%%%%%%%%%%%%%%%%%%%%%%%%%
\section{Energy-Momentum Complexes}
The conservation laws of energy and momentum  for an isolated, i.e., no external force acting on the system,
physical system in the Special Theory of Relativity are expressed by a set of differential equations. Defining
$T^{\mu}_{\nu}$ as the symmetric energy-momentum tensor of matter and of all non-gravitational fields, the
conservation laws are given by
\be
T^{\mu}_{\nu,\, \mu} \equiv \frac{\partial T^{\mu}_{\nu}}{\partial
x^{\mu}}=0\ee where \be \rho=T^{t}_{t}\hspace{1cm}j^{i}=T^{i}_{t}\hspace{1cm}p_{i}=-T^{t}_{i}
\ee
are the energy density, the energy current density, the momentum density, respectively, and Greek indices run over the spacetime
labels while Latin indices run over the spatial coordinate values.
\par\noindent
Making the transition from Special to General Theory of Relativity, one adopts a simplicity principle which is
called principle of minimal gravitational coupling. As a result of this, the conservation equation is now written
as
\be
T^{\mu}_{\nu;\, \mu} \equiv \frac{1}{\sqrt{-g}}\frac{\partial}{\partial
x^{\mu}}\left(\sqrt{-g}\,T^{\mu}_{\nu}\right)-\Gamma^{\kappa}_{\nu\lambda}T^{\lambda}_{\kappa}=0
\ee
where $g$ is the determinant of the metric tensor $g_{\mu\nu}(x)$.
The conservation equation may also be written as
\be
\frac{\partial}{\partial x^{\mu}}\left(\sqrt{-g}\,T^{\mu}_{\nu}\right)=\xi_{\nu} \ee where \be
\xi_{\nu}=\sqrt{-g}\Gamma^{\kappa}_{\nu\lambda}T^{\lambda}_{\kappa}
\ee
is a non-tensorial object. For $\nu=t$
this means that the matter energy is not a conserved quantity for the physical system\footnote{It is possible to
restore the conservation law by introducing a local inertial system for which at a specific spacetime point
$\xi_{\nu}=0$ but this equality by no means holds in general.}. From a physical point of view, this lack of energy
conservation can be understood as the possibility of transforming matter energy into gravitational energy and vice
versa. However, this remains a problem and it is widely believed that in order to be solved one has to take into
account the gravitational energy.

\par\noindent
By a well-known procedure, the non-tensorial object $\xi_{\nu}$ can be written  as
\be
\xi_{\nu}=-\frac{\partial}{\partial x^{\mu}}\left(\sqrt{-g}\,\vartheta^{\mu}_{\nu}\right)
\ee
where $\vartheta^{\mu}_{\nu}$ are certain functions of the metric tensor and its first order derivatives.
Therefore, the energy-momentum tensor of matter $T^{\mu}_{\nu}$ is replaced by the expression
\be
\theta^{\mu}_{\nu}=\sqrt{-g}\left(T^{\mu}_{\nu}+\vartheta^{\mu}_{\nu}\right)
\ee
which is called energy-momentum complex since it is a combination of the tensor
$T^{\mu}_{\nu}$ and a pseudotensor $\vartheta^{\mu}_{\nu}$ which
describes the energy and  momentum of the gravitational field. The energy-momentum complex satisfies a
conservation law in the ordinary sense, i.e.,
\be \theta^{\mu}_{\nu,\, \mu}=0
\ee
and it can be written as
\be
\theta^{\mu}_{\nu}=\chi^{\mu\lambda}_{\nu\,\,\,,\lambda}
\ee
where $\chi^{\mu\lambda}_{\nu}$ are called
superpotentials and are functions of the metric tensor and its first order derivatives.

\par\noindent
It is evident that the energy-momentum complex is not uniquely determined by the condition that its usual
divergence is zero since it can always been added to the energy-momentum complex a quantity with an identically
vanishing divergence.
%%%%%%%%%%%%%%%%%%%%%%%%%%%%%%%%%%%%%%%%%%%%%%%%%%%%%%%%%%%%%%%%%%%%%%%%%%%%%%%%%%%%%%%%%%%%%%%%%
%%%%%%%%%%%%%%%%%%%%%%%%%%%%%%%%%%%%%%%%%%%%%%%%%%%%%%%%%%%%%%%%%%%%%%%%%%%%%%%%%%%%%%%%%%%%%%%%%
%%%%%%%%%%%%%%%%%%%%%%%%%%%%%%%%%%%%%%%%%%%%%%%%%%%%%%%%%%%%%%%%%%%%%%%%%%%%%%%%%%%%%%%%%%%%%%%%%
%%%%%%%%%%%%%%%%%%%%%%%%%%%%%%%%%%%%%%%%%%%%%%%%%%%%%%%%%%%%%%%%%%%%%%%%%%%%%%%%%%%%%%%%%%%%%%%%%
\section{BTZ black hole}
In 1992 Ba\~{n}ados, Teitelboim, and Zanelli discovered a black hole solution (known as  BTZ black hole) in
($2+1$) dimensions  \cite{jorge}. Till that time it was believed that no black hole solution exists in
three-dimensional spacetimes \cite{abbott}. Ba\~{n}ados, Teitelboim, and Zanelli found a vacuum solution to
Einstein gravity with a negative cosmological constant.

\par\noindent
The starting point was the action in a three-dimensional
theory of gravity
\be
S=\int d^{3}x\sqrt{-g}\left(R+\frac{2}{l^{2}}\right)
\label{action}
\ee
where the radius of curvature $l$ is related to the cosmological constant by $\Lambda=-l^{-2}$.

\par\noindent
It is straightforward to check that the Einstein's field equations
\be
R_{\mu\nu}-\frac{1}{2}g_{\mu\nu}\left(R+ \frac{2}{l^{2}}\right) = 0
\label{fieldeqns}
\ee
are solved by the metric\footnote{The form of the BTZ metric in quasi-Cartesian coordinates can
be obtained by making the transformations \bea
x&=&r\cos (\phi)\nn\\
y&=&r\sin(\phi)\nn
\hspace{1ex}.
\eea}
\be
ds^{2}=N^{2}(r)dt^{2} -\frac{dr^{2}}{N^{2}(r)}-r^{2}\left(N^{\phi}(r)dt+d\phi\right)^{2},
\label{metric}
\ee
where the squared lapse $N^{2}(r)$ and the angular shift $N^{\phi}(r)$ are given by
\be
N^{2}(r)=\frac{r^2}{l^2}- M +\frac{J^{2}}{4r^{2}}\hspace{1ex},\hspace{4ex} N^{\phi}(r)=-\frac{J}{2r^{2}}
\label{lapse}
\ee
with $-\infty<t<+\infty$, $0<r<+\infty$, and $0\leq\phi\leq 2\pi$.
Since the metric (\ref{metric}) satisfies the Einstein's field equations with a negative
cosmological constant (see (\ref{fieldeqns})), the metric is locally Anti-de Sitter
\be
ds^{2}=\left(1+ \frac{r^2}{l^2}\right)dt^{2}-\frac{dr^{2}}{\left(1+\displaystyle{\frac{r^2}{l^2}}\right)}
-r^{2}d\phi ^{2}
\ee
and it can only differ from Anti-de Sitter space in its global properties.
The two constants $M$ and $J$ are the conserved quantities: mass and angular momentum, respectively.
The lapse function $N(r)$ vanishes for two values of the radial coordinate $r$ given by
\be
r^{2}_{\pm}=\frac{l^2}{2}\left[M\pm \sqrt{M^2 - \left(\frac{J}{l}\right)^2}\right]
\hspace{1ex}.
\ee
The largest root, $r_{+}$, is the black hole horizon. It is evident that in order for the
horizon to exist one must have
\be
M>0 \hspace{1ex},|J|\leq M l
\hspace{1ex}.
\ee
Therefore, negative black hole masses are excluded from the physical spectrum.
There is, however, an important exceptional case. When one sets $M=-1$ and $J=0$, the singularity, i.e., $r=0$,
disappears. There is neither a horizon nor a singularity to hide. The configuration is again
that of Anti-de Sitter space. Thus, Anti-de Sitter emerges as a ``bound state'', separated from the continuous
black hole spectrum by a mass gap of one unit.
\par\noindent
For the specific case of spinless ($J=0$) BTZ black hole, the line element (\ref{metric}) takes the simple form
\be
ds^{2}=\left(\frac{r^{2}}{l^{2}}- M\right)dt^{2} -\frac{dr^{2}}{\left(\displaystyle{\frac{r^{2}}{l^{2}}}- M\right)}
-r^{2}d\phi^{2}
\hspace{1ex}.
\label{spinless}
\ee
As it is stated in the Introduction, metric (\ref{metric}) of the rotating ($2+1$)-dimensional BTZ black hole
is not asymptotically (that is as $r\rightarrow \infty$) flat
\be
ds^{2}=dt^2 -dr^2 -d\phi^2
\hspace{1ex}.
\ee
Specifically,  the BTZ black hole metric (\ref{metric}) for large $r$ becomes
\be
ds^{2}=\left(\frac{r}{l}\right)^{2}dt^{2}-\left(\frac{r}{l}\right)^{-2}dr^2 -r^{2}d\phi ^{2}
\ee
which coincides with  the asymptotic form of Anti-de Sitter metric.
\par\noindent
Additionally, the singularity at $r=0$ is much
weaker than that of the Schwarzschild spacetime.  The curvature scalars for the case of
Schwarzschild black hole are of the form
\be
R^{\alpha\beta\gamma\delta}R_{\alpha\beta\gamma\delta}\sim \frac{M^{2}}{r^{6}}
\hspace{1ex}.
\ee
It is easily seen that the singularity of Schwarzschild black hole at $r=0$ is manifested
by the power-law divergence of curvature scalars. For the case of BTZ black hole, the curvature scalars
are
\be
R^{\alpha\beta\gamma\delta}R_{\alpha\beta\gamma\delta}\sim \frac{1}{l^4}
\ee
and thus the BTZ solution has at most a $\delta$-function singularity at $r=0$,
since everywhere else
the spacetime is of constant curvature \cite{singularity}.
%%%%%%%%%%%%%%%%%%%%%%%%%%%%%%%%%%%%%%%%%%%%%%%%%%%%%%%%%%%%%%%%%%%%%%%%%%%%%%%%%%%%%%%%%%%%%%%%%
%%%%%%%%%%%%%%%%%%%%%%%%%%%%%%%%%%%%%%%%%%%%%%%%%%%%%%%%%%%%%%%%%%%%%%%%%%%%%%%%%%%%%%%%%%%%%%%%%
%%%%%%%%%%%%%%%%%%%%%%%%%%%%%%%%%%%%%%%%%%%%%%%%%%%%%%%%%%%%%%%%%%%%%%%%%%%%%%%%%%%%%%%%%%%%%%%%%
%%%%%%%%%%%%%%%%%%%%%%%%%%%%%%%%%%%%%%%%%%%%%%%%%%%%%%%%%%%%%%%%%%%%%%%%%%%%%%%%%%%%%%%%%%%%%%%%%
\section{Landau-Lifshitz's Prescription}
For a three-dimensional background the energy-momentum complex of Landau-Lifshitz \cite{ll} is defined by
\be
L^{\mu\nu}=\frac{1}{2\kappa}S^{\mu\kappa\nu\lambda}_{\hspace{4ex},\kappa\lambda}
\label{lltheta}
\ee
where $\kappa$ is the gravitational coupling constant.
The Landau-Lifshitz's superpotential $S^{\mu\kappa\nu\lambda}$ is
\be
S^{\mu\kappa\nu\lambda}=-g\left(g^{\mu\nu}g^{\kappa\lambda}-g^{\mu\lambda}g^{\kappa\nu}\right)
\hspace{1ex}.
\label{llsuperpot}
\ee
The  energy-momentum complex of Landau-Lifshitz is symmetric in its indices
\be
L^{\mu\nu}=L^{\nu\mu}
\hspace{1ex}.
\ee
Energy and momentum in Landau-Lifshitz's prescription for a
three-dimensional background are given by
\be
P^{\mu}=\int\int L^{t \mu}dx^{1}dx^{2}
\hspace{1ex}.
\label{llmomentum}
\ee
In particular, the energy of a physical system in a three-dimensional background takes the
form
\be
E=\int\int L^{tt}dx^{1}dx^{2}
\hspace{1ex}.
\label{llenergy}
\ee
It is important to underscore that all
calculations in the Landau-Lifshitz's prescription have to be restricted to the use of quasi-Cartesian
coordinates.

\par\noindent
Our primary task is to explicitly compute the Landau-Lifshitz's superpotentials (\ref{llsuperpot}). For the
($2+1$)-dimensional rotating BTZ black hole (\ref{metric}) there are only thirty six  non-zero Landau-Lifshitz's
superpotentials (see Appendix). Substituting Landau-Lifshitz's superpotentials into equation (\ref{lltheta}),
one gets the energy density distribution
\be
L^{tt}=\frac{1}{\kappa r^{2}}\left(\frac{r^{2}}{l^2}-\displaystyle{\frac{J^{2}}{4r^{2}}}\right)
\left(\frac{r^{2}}{l^2}-M+\displaystyle{\frac{J^{2}}{4r^{2}}}\right)^{-2}
\hspace{1ex}.
\label{lldensity}
\ee
Therefore, if we substitute the expression of the energy density (\ref{lldensity})
into equation (\ref{llenergy}), we get
\bea
E_{ext}&=&2\pi \int^{\infty}_{r_{0}}L^{tt}rdr\nn\\
&=&\frac{\pi}{\kappa}\left(\frac{r_{0}^{2}}{l^2}-M+\displaystyle{\frac{J^{2}}{4r^{2}_{0}}}\right)^{-1}
\hspace{1ex}.
\eea
If we set the asymptotic value of the total gravitational mass of the system to be $M(\infty)$, then it is
clear that the energy distribution associated with the rotating ($2+1$)-dimensional BTZ black hole under study,
which is contained in a ``one-sphere''  of radius $r_{0}$, will be
\be
E(r_{0})=M(\infty)-\frac{\pi}{\kappa}\left(\frac{r_{0}^{2}}{l^2}-M+
\displaystyle{\frac{J^{2}}{4r^{2}_{0}}}\right)^{-1}
\hspace{1ex}.
\label{LLenergy}
\ee
A couple of comments are in order.
Firstly, it is easily seen that for the case of the black hole horizon, i.e., $r_{0}=r_{+}$, the energy
distribution given by equation (\ref{LLenergy}) becomes infinite.
The infiniteness of  the energy distribution stems from the infiniteness of the energy density
given by equation (\ref{lldensity}) when we set $r$ equal to $r_{+}$.
However, this value for the radial coordinate should be excluded.
It is known that it is meaningless the energy density of a closed system to be infinite and
this is the case here since the energy density (\ref{lldensity}) is restricted to a ``one-sphere''  of radius $r_{0}$.
Secondly, a neutral test particle experiences at a finite distance $r_{0}$
the gravitational field of the effective gravitational mass\footnote{Cohen and Gautreau were the first
who gave a definition of effective mass by implementing Whittaker's theorem \cite{effectivemass1}. They derived
the effective mass for a Reissner-Nordstr$\ddot{o}$m black hole. Cohen and de Felice computed the total effective
gravitational mass that a particle experiences while approaching the source of a Kerr-Newman spacetime
\cite{effectivemass2}.
Unfortunately, the expression for the total effective mass of the Kerr-Newman spacetime didn't incorporate the
rotational contribution (setting the electric charge equal to zero gives the total mass $M$).
R. Kulkarni, V. Chellathurai, and N. Dadhich generalized the result
of Cohen and de Felice. They defined the total effective gravitational mass for
a Kerr spacetime where the rotational effects were incorporated \cite{effectivemass3}.}
described by expression (\ref{LLenergy}).
Thirdly, the  energy-momentum complex of Landau-Lifshitz as formulated
here for the rotating ($2+1$)-dimensional BTZ black hole satisfies the local conservation laws
\be
\frac{\partial}{\partial x^{\nu}}\,L^{\mu\nu}=0
\hspace{1ex}.
\ee
%%%%%%%%%%%%%%%%%%%%%%%%%%%%%%%%%%%%%%%%%%%%%%%%%%%%%%%%%%%%%%%%%%%%%%%%%%%%%%%%%%%%%%%%%%%%%%%%%
%%%%%%%%%%%%%%%%%%%%%%%%%%%%%%%%%%%%%%%%%%%%%%%%%%%%%%%%%%%%%%%%%%%%%%%%%%%%%%%%%%%%%%%%%%%%%%%%%
%%%%%%%%%%%%%%%%%%%%%%%%%%%%%%%%%%%%%%%%%%%%%%%%%%%%%%%%%%%%%%%%%%%%%%%%%%%%%%%%%%%%%%%%%%%%%%%%%
%%%%%%%%%%%%%%%%%%%%%%%%%%%%%%%%%%%%%%%%%%%%%%%%%%%%%%%%%%%%%%%%%%%%%%%%%%%%%%%%%%%%%%%%%%%%%%%%%
%%%%%%%%%%%%%%%%%%%%%%%%%%%%%%%%%%%%%%%%%%%%%%%%%%%%%%%%%%%%%%%%%%%%%%%%%%%%%%%%%%%%%%%%%%%%%%%%%
\section{Weinberg's Prescription}
For a three-dimensional background the energy-momentum complex of Weinberg \cite{weinberg} is defined by
\be
\tau^{\nu\lambda}=\frac{1}{2\kappa}Q^{\rho\nu\lambda}_{\hspace{3ex},\rho}
\label{weintheta}
\ee
where Weinberg's superpotential $Q^{\rho\nu\lambda}$ is given by
\be
Q^{\rho\nu\lambda}=\frac{\partial h^{\mu}_{\mu}}{\partial
x_{\nu}}\eta^{\rho\lambda}- \frac{\partial h^{\mu}_{\mu}}{\partial x_{\rho}}\eta^{\nu\lambda}- \frac{\partial
h^{\mu\nu}}{\partial x^{\mu}}\eta^{\rho\lambda}+ \frac{\partial h^{\mu\rho}}{\partial x^{\mu}}\eta^{\nu\lambda}+
\frac{\partial h^{\nu\lambda}}{\partial x_{\rho}}- \frac{\partial h^{\rho\lambda}}{\partial x_{\nu}}
\hspace{1ex}.
\label{weinsuperpot}
\ee
The symmetric tensor $h_{\mu\nu}$ reads
\be
h_{\mu\nu}=g_{\mu\nu}-\eta_{\mu\nu}
\ee
and
$\eta^{\mu\nu}$ is the Minkowskian metric
\be
(\eta^{\mu\nu})=\left[
\begin{array}{rrr}
1&0&0\\
0&-1&0\\
0&0&-1\\
\end{array}\right]
\label{minkowski}
\hspace{1ex}.
\ee
It is important to stress that all indices on $h_{\mu\nu}$ and/or
$\partial/\partial x_{\mu}$ are raised, or lowered, with the use of the Minkowskian metric. Additionally, the
energy-momentum complex of Weinberg is symmetric in its indices
\be
\tau^{\mu\nu}=\tau^{\nu\mu}
\ee
while the superpotential $Q^{\rho\nu\lambda}$ is antisymmetric in its first two indices
\be
Q^{\rho\nu\lambda}=-Q^{\nu\rho\lambda}
\hspace{1ex}.
\ee
Energy and momentum in Weinberg's prescription for a
three-dimensional background are given by
\be
P^{\nu}=\int\int \tau^{\nu t}dx^{1}dx^{2}
\label{weinmomentum}
\hspace{1ex}.
\ee
Specifically, the energy of a physical system in a three-dimensional background is
\be
E=\int\int \tau^{tt}dx^{1}dx^{2}
\hspace{1ex}.
\label{weinenergy}
\ee
It should be underscored again that all
calculations in the Weinberg's prescription have to be performed using quasi-Cartesian coordinates.

\par\noindent
Since our aim is to evaluate the energy distribution associated with the rotating ($2+1$)-dimensional BTZ black
hole background described by the line element (\ref{metric}), we firstly evaluate the Weinberg's superpotentials.
There are sixteen  non-zero superpotentials in the Weinberg's prescription (see Appendix). Substituting
Weinberg's superpotentials  into equation (\ref{weintheta}), the energy density distribution takes the form
\be
\tau^{tt}=\frac{1}{\kappa r^{2}}\left(\frac{r^{2}}{l^2}-\displaystyle{\frac{J^{2}}{4r^{2}}}\right)
\left(\frac{r^{2}}{l^2}-M+\displaystyle{\frac{J^{2}}{4r^{2}}}\right)^{-2}
\label{weindensity}
\ee

\par\noindent
It is evident that we have derived the same energy density distribution
associated with the rotating ($2+1$)-dimensional BTZ black hole
as in the case of the Landau-Lifshitz's prescription (see equation (\ref{lldensity})).
Thus, the energy distribution
will be exactly the same with the one computed in Landau-Lifshitz's prescription (see equation (\ref{LLenergy})).
The comments concerning the neutral test particle, the
effective gravitational mass, and the local conservation laws made in the preceding section also hold here.

%%%%%%%%%%%%%%%%%%%%%%%%%%%%%%%%%%%%%%%%%%%%%%%%%%%%%%%%%%%%%%%%%%%%%%%%%%%%%%%%%%%%%%%%%%%%%%%%%
%%%%%%%%%%%%%%%%%%%%%%%%%%%%%%%%%%%%%%%%%%%%%%%%%%%%%%%%%%%%%%%%%%%%%%%%%%%%%%%%%%%%%%%%%%%%%%%%%
%%%%%%%%%%%%%%%%%%%%%%%%%%%%%%%%%%%%%%%%%%%%%%%%%%%%%%%%%%%%%%%%%%%%%%%%%%%%%%%%%%%%%%%%%%%%%%%%%
%%%%%%%%%%%%%%%%%%%%%%%%%%%%%%%%%%%%%%%%%%%%%%%%%%%%%%%%%%%%%%%%%%%%%%%%%%%%%%%%%%%%%%%%%%%%%%%%%
\section{Conclusions}
In this work we have explicitly evaluated the energy distribution, contained in a ``one-sphere" of fixed radius
$r_{0}$, of a rotating ($2+1$)-dimensional BTZ black hole. The gravitational background under consideration is an
exact vacuum solution of Einstein's field equations in the presence of a negative cosmological constant. It is
characterized by two conserved charges: the mass $M$ and the angular momentum $J$. The energy distribution is
obtained using two different energy-momentum complexes, specifically these are the energy-momentum complexes of
Landau-Lifshitz and Weinberg. Both prescriptions give exactly the same energy distribution for the specific
gravitational background.
Consequently, the results obtained here support the claim\footnote{The claim that different pseudotensors
give same results for local quantities was first stated by Virbhadra \cite{virbhadra} and later on by
Aguirregabiria, Chamorro, and Virbhadra in \cite{virbhadra1}.
The author provided evidence in support of this claim for the case of a non-static spinless
($2+1$)-dimensional black hole with an outflux of null radiation \cite{claim}. However, this is not the case
for the two-dimensional stringy black hole backgrounds. In particular, it was shown that M{\o}ller's
energy-momentum complex provides meaningful physical results while the Einstein's
energy-momentum complex fails to do so \cite{noclaim}.}
that for a given gravitational background, different energy-momentum complexes can give exactly the same energy and momentum distributions in
three dimensions as they do in four dimensions. However, it should be stress that
both Landau-Lifshitz's and Weinberg's energy-momentum complexes are symmetric
(on the contrary Einstein's energy-momemtum complex is not symmetric)
and do not depend on the second derivative of the metric.
\par
It should also be pointed out that the energy distribution derived here can be regarded as the effective
gravitational mass experienced by a neutral test particle placed in the rotating ($2+1$)-dimensional black hole
background under consideration.
\par
Setting the angular momentum $J$ equals to zero, one can derive the energy distribution associated with the
spinless ($2+1$)-dimensional BTZ black hole. Recently, I.-C. Yang and I. Radinschi \cite{irina} calculated the energy
distributions associated with four ($2+1$)-dimensional black hole solutions by utilizing Einstein's
and M{\o}ller's energy-momentum complexes. One of these spacetimes is the
spinless ($2+1$)-dimensional BTZ black hole (or, the uncharged black hole solution as named in their manuscript).
I.-C. Yang and I. Radinschi showed that the energy distribution of Einstein's energy-momentum complex is different from
the one of Moller's energy-momentum complex.
Furthermore, both expressions for the energy distribution associated with the
spinless ($2+1$)-dimensional BTZ black hole are different from the one we
have derived in the present analysis.
\par
Finally, since the (2+1)-dimensional BTZ black hole is an asymptotically Anti-de-Sitter spacetime (AAdS), it would
be an oversight not to mention that in the last years due to the AdS/CFT correspondence there has been much
progress in obtaining finite stress energy tensors of AAdS spacetimes\footnote{For a short review see
\cite{sken}}. The gravitational stress energy tensor is in general infinite due to the infinite volume of the
spacetime. In order to find a meaningful definition of gravitational energy one should subtract the divergences.
The proposed prescriptions so far were ad hoc in the sense that one has to embed the boundary in some reference
spacetime. The important drawback of this method is that it is not always possible to find the suitable reference
spacetime. Skenderis and collaborators\footnote{Right after the first work of  Henningson and Skenderis
\cite{hen1}, Nojiri and Odintsov \cite{nojiri1} calculated a finite gravitational stress energy tensor for an AAdS
space where the dual conformal field theory is dilaton coupled. Furthermore, Nojiri and Odintson \cite{nojiri2},
and Ogushi \cite{nojiri3} found well-defined gravitational stress energy tensors for AAdS spacetimes in the
framework of higher derivative gravity and of gauged supergravity with single dilaton respectively. }
\cite{hen1,hen2,skenderis,haro}, and also Balasubramanian and Kraus \cite{bal}, described and implemented a new
method  which provides an intrinsic definition of the gravitational stress energy tensor. The computations are
universal in the sense that apply to all AAdS spacetimes. Therefore, it is nowadays right to state that the issue
of the gravitational stress energy tensor for any AAdS spacetime has been thoroughly understood.
%%%%%%%%%%%%%%%%%%%%%%%%%%%%%%%%%%%%%%%%%%%%%%%%%%%%%%%%%%%%%%%%%%%%%%%%%%%%%%%%%%%%%%%%%%%%%%%%%%%%%%
%%%%%%%%%%%%%%%%%%%%%%%%%%%%%%%%%%%%%%%%%%%%%%%%%%%%%%%%%%%%%%%%%%%%%%%%%%%%%%%%%%%%%%%%%%%%%%%%%%%%%%
%%%%%%%%%%%%%%%%%%%%%%%%%%%%%%%%%%%%%%%%%%%%%%%%%%%%%%%%%%%%%%%%%%%%%%%%%%%%%%%%%%%%%%%%%%%%%%%%%%%%%%
\section*{Acknowledgements}
The author is indebted to K.S. Virbhadra for useful
suggestions and comments on the manuscript.
The author is also grateful to K. Skenderis for his enlightening comments on holographic renormalization method.
The author has been supported by the European Research and Training Network ``EUROGRID-Discrete Random Geometries:
from Solid State Physics to Quantum Gravity" (HPRN-CT-1999-00161).
%%%%%%%%%%%%%%%%%%%%%%%%%%%%%%%%%%%%%%%%%%%%%%%%%%%%%%%%%%%%%%%%%%%%%%%%%%%%%%%%%%%%%%%%%%%%%%%%%%%%%%
%%%%%%%%%%%%%%%%%%%%%%%%%%%%%%%%%%%%%%%%%%%%%%%%%%%%%%%%%%%%%%%%%%%%%%%%%%%%%%%%%%%%%%%%%%%%%%%%%%%%%%
%%%%%%%%%%%%%%%%%%%%%%%%%%%%%%%%%%%%%%%%%%%%%%%%%%%%%%%%%%%%%%%%%%%%%%%%%%%%%%%%%%%%%%%%%%%%%%%%%%%%%%
%%%%%%%%%%%%%%%%%%%%%%%%%%%%%%%%%%%%%%%%%%%%%%%%%%%%%%%%%%%%%%%%%%%%%%%%%%%%%%%%%%%%%%%%%%%%%%%%%%%%%%
%%%%%%%%%%%%%%%%%%%%%%%%%%%%%%%%%%%%%%%%%%%%%%%%%%%%%%%%%%%%%%%%%%%%%%%%%%%%%%%%%%%%%%%%%%%%%%%%%%%%%%
\section*{Appendix}
\subsection*{Landau-Lifshitz's Superpotentials}
Using equation (\ref{llsuperpot}) we can explicitly evaluate the superpotentials in the Landau-Lifshitz's
prescription. There are thirty six non-vanishing superpotentials but for brevity we give the four superpotentials
which are necessary to compute the energy density distribution (\ref{lldensity})
\bea
S^{txtx}&=&\frac{J^{2}x^{2}+4\left(x^{2}+y^{2}\right)\left(-Mx^{2}+y^{2}+l^{-2} x^{4}+l^{-2} x^{2}y^{2}\right)}
{\left(x^{2}+y^{2}\right)\left(J^{2}+4\left(x^{2}+y^{2}\right)
\left(-M+l^{-2}\left(x^{2}+y^{2}\right)\right)\right)}\,,\nn\\
S^{txty}&=&\frac{xy\left(J^{2}+4\left(x^{2}+y^{2}\right)\left(-1-M+l^{-2}\left(x^{2}+y^{2}\right)\right)\right)}
{\left(x^{2}+y^{2}\right)\left(J^{2}+4\left(x^{2}+y^{2}\right)
\left(-M+l^{-2}\left(x^{2}+y^{2}\right)\right)\right)}\,,\nn\\
S^{tytx}&=&\frac{xy\left(J^{2}+4\left(x^{2}+y^{2}\right)\left(-1-M+l^{-2}\left(x^{2}+y^{2}\right)\right)\right)}
{\left(x^{2}+y^{2}\right)\left(J^{2}+4\left(x^{2}+y^{2}\right)
\left(-M+l^{-2}\left(x^{2}+y^{2}\right)\right)\right)}\,,\nn\\
S^{tyty}&=&\frac{4x^{4}\left(1+l^{-2} y^{2}\right)+4x^{2}y^{2}\left(1-M+2l^{-2} y^{2}\right)+
y^{2}\left(J^{2}-4My^{2}+4l^{-2} y^{4}\right)} {\left(x^{2}+y^{2}\right)\left(J^{2}+4\left(x^{2}+y^{2}\right)
\left(-M+l^{-2}\left(x^{2}+y^{2}\right)\right)\right)}\,
\hspace{1ex}.\nn \eea
%%%%%%%%%%%%%%%%%%%%%%%%%%%%%%%%%%%%%%%%%%%%%%%%%%%%%%%
\subsection*{Weinberg's Superpotentials}
Using equation (\ref{weinsuperpot}) we can explicitly evaluate the superpotentials in Weinberg's prescription.
There are sixteen non-vanishing superpotentials but for brevity we give the four superpotentials
which are necessary to compute the energy density distribution (\ref{weindensity}).

\bea
Q^{txt}&=&-\frac{x\left(J^{2}+4\left(x^{2}+y^{2}\right)\left(-1-M+l^{-2} \left(x^{2}+y^{2}\right)\right)\right)}
{\left(x^{2}+y^{2}\right)\left(J^{2}+4\left(x^{2}+y^{2}\right)\left(-M+l^{-2}
\left(x^{2}+y^{2}\right)\right)\right)}\,,\nn\\
Q^{tyt}&=&-\frac{y\left(J^{2}+4\left(x^{2}+y^{2}\right)\left(-1-M+l^{-2} \left(x^{2}+y^{2}\right)\right)\right)}
{\left(x^{2}+y^{2}\right)\left(J^{2}+4\left(x^{2}+y^{2}\right)\left(-M+l^{-2}
\left(x^{2}+y^{2}\right)\right)\right)}\,,\nn\\
Q^{xtt}&=&\frac{x\left(J^{2}+4\left(x^{2}+y^{2}\right)\left(-1-M+l^{-2} \left(x^{2}+y^{2}\right)\right)\right)}
{\left(x^{2}+y^{2}\right)\left(J^{2}+4\left(x^{2}+y^{2}\right)\left(-M+l^{-2}
\left(x^{2}+y^{2}\right)\right)\right)}\,,\nn\\
Q^{ytt}&=&\frac{y\left(J^{2}+4\left(x^{2}+y^{2}\right)\left(-1-M+l^{-2} \left(x^{2}+y^{2}\right)\right)\right)}
{\left(x^{2}+y^{2}\right)\left(J^{2}+4\left(x^{2}+y^{2}\right)\left(-M+l^{-2}
\left(x^{2}+y^{2}\right)\right)\right)}\,
\hspace{1ex}.\nn
\eea
%%%%%%%%%%%%%%%%%%%%%%%%%%%%%%%%%%%%%%%%%%%%%%%%%%%%%%%%%%%%%%%%%%%%%%%%%%%%%%%%%%%%%%%%%%%%%%%%%%%%%%
%%%%%%%%%%%%%%%%%%%%%%%%%%%%%%%%%%%%%%%%%%%%%%%%%%%%%%%%%%%%%%%%%%%%%%%%%%%%%%%%%%%%%%%%%%%%%%%%%%%%%%
%%%%%%%%%%%%%%%%%%%%%%%%%%%%%%% BIBLIOGRAPHY %%%%%%%%%%%%%%%%%%%%%%%%%%%%%%%%%%%%%%%%%%%%%%%%%%%%%%%%%

%%%%%%%%%%%%%%%%%%%%%%%%%%%%%%%%%%%%%%%%%%%%%%5
\end{document}